\documentclass[a4paper,12pt]{article}
\usepackage[english]{babel}
\usepackage{mathrsfs}
\usepackage{amsmath}
\usepackage{amsfonts}
\usepackage{amssymb}
\usepackage{latexsym}
\newcommand{\bd}{\begin{definition}}
\newcommand{\ed}{\end{definition}}

\newcommand{\bt}{\begin{theorem}}
\newcommand{\et}{\end{theorem}}
\newcommand{\bi}{\begin{itemize}}
\newcommand{\ei}{\end{itemize}}
\newcommand{\ben}{\begin{enumerate}}
\newcommand{\een}{\end{enumerate}}
\newcommand{\al}{\mbox{$ \alpha  $}}
\newcommand{\be}{\mbox{$ \beta  $}}

\newcommand{\Ga}{\mbox{$ \Gamma(\alpha_1,...,\alpha_n)  $}}
\newcommand{\Gb}{\mbox{$ \Gamma(\beta_1,...,\beta_n)  $}}

\newcommand{\hc}{\mbox{$ \mathcal {H} $}}

\newcommand{\ag}{\mbox{$ \mathcal {A}/ \mathcal {G} $}}
\newcommand{\bag}{\mbox{$ \overline {\mathcal {A}/ \mathcal {G}} $}}
\newcommand{\ibag}{\mbox{$ \int_{\overline {\mathcal {A}/ \mathcal {G}}} \, $}}

\newcommand{\cs}{\mbox{$ C^{\ast} $-}}

\newcommand{\tl}{\mbox{$ {Loop}_{\star}(M) $}}
\newcommand{\pt}{\mbox{$ Path(M)$}}
\newtheorem{definition}{Definition}[section]
\newtheorem{theorem}{Theorem}[section]

\newtheorem{proposition}{Proposition}[section]

\newcommand{\naturali}{{\hbox{{\rm N}\kern-.2em\hbox{\rm R}}}}
\newcommand{\interi}{{\ \hbox{{\rm Z}\kern-.6em\hbox{\bf Z}}}}
\newcommand{\reali}{{\hbox{{\rm I}\kern-.2em\hbox{\rm R}}}}
\newcommand{\complessi}{{\ \hbox{{\rm I}\kern-.6em\hbox{\bf C}}}}

\begin{document}
\begin{center}

{\Large{\bf Inductive Construction of the Loop Transform 

for Abelian Gauge Theories}}
\vskip 1cm
{\bf Maria Cristina Abbati$^{*}$, Alessandro Mani\`a$^{*}$ and
Edoardo Provenzi$^{**}$}
\end{center}
\vskip 0,2cm
\noindent{(*) Dipartimento di Fisica, Universit\`a degli Studi di Milano
 and  Istituto Nazionale di Fisica Nucleare, Sezione di Milano, Milano, Italy\\
e-mail: maria.cristina.abbati@mi.infn.it, alessandro.mania@mi.infn.it}

\noindent{(**) Dipartimento di Matematica, Universit\`a di Genova, Genova, Italy}

\vskip 0,5cm
\begin{center}{\bf Abstract}
\end{center}
{We construct the loop transform
in the case of Abelian gauge theories as a unitary operator given by the inductive limit of Fourier transforms on
tori. We also show that its range, i.e. the space of kinematical states of the quantum loop representation, is the Hilbert space of square integrable complex valued functions on the group of hoops.}

\vskip 0,2cm
\noindent{\bf Mathematical Subject Classification}:  22E67, 81T13
\vskip 0,2cm
\noindent{\bf Key words:}{ loop transform, projective limit of groups}

\section{Introduction}

The history of the loop transform begins in 1990 when Rovelli and
Smolin \cite{rs}  proposed, in the context of canonical quantum
gravity,
a formal transform  to pass from
functions   on the space of connections  to   functions
 of loops.
In a gauge theory where the gauge group $G$ is assumed to be a closed
subgroup of $U(N)$, the loop transform $\ell_\psi$ of a function
$\psi$ of connections
is given by
$$ \ell_\psi(\al):=\int_{\cal A}
\overline{T_\alpha(A)}\psi(A){\cal D}A
$$ where $\cal A$ is the  space  of smooth principal connections
of a fixed principal fiber bundle $P(M,G)$, ${\cal D}A$ is a formal measure on $\cal A$,  $\al$ is a loop in $M$ at least piecewise $C^1$ and  $T_\alpha$ is the associated Wilson function, which relates a connection $A\in{\cal A}$ to the normalized trace of the holonomy of $A$ around $\alpha$, usually written
$$T_\alpha(A)=\frac{1}{N} \, Tr({\mathcal P}\exp \oint_{\al}A) \;, $$
where $\cal P$ denotes the parallel transport.

\vskip 0,2cm
The role of this transform is analogous to that of the Fourier
transform in quantum mechanics, which enables to pass from the position representation
to the momentum representation.
In the canonical quantization of gauge theories,
 the loop transform should relate the {\em connection representation},
in which states  are  functions on the configuration space
 $\cal A$, to the {\em loop representation} in which states are
 functions on the loop space.
Formally, the Wilson functions $T_{\alpha}$ play the r{o}le of the
 phase factors in the Fourier transforms and $\cal A$ the r{o}le of
 the finite  dimensional configuration space ${\mathbb R}^n$ in quantum mechanics.

The loop transform is relevant for gauge theories invariant under diffeomorphisms as the 
Euclidean  formulation of General Relativity or the Chern-Simon theory in three dimensions. In fact, if $\psi$ and
 ${\cal D}A$ are invariant under   diffeomorphisms and gauge
transformations, then the function $\ell_\psi$ is a topological
invariant of the manifold $M$ and it naturally satisfies the gauge and
diffeomorphism constraints in the quantum theory \cite{bru} .

The connection and the  loop representations are equivalent if and
only if the loop transform is an unitary operator. 
 The first problem to get a loop transform was the
construction of the measure ${\cal D}A$. Measures invariant under
gauge transformations and diffeomorphisms were obtained by
Ashtekar and Lewandowski \cite{asle} and by Baez \cite{baez1} on a
suitable compact space $\bag$ containing densely $\ag$, the space
of connections modulo gauge transformations. Actually, $\bag$ is
the spectrum of the $C^*$-algebra generated by the Wilson
functions $T_{\alpha}$ in the case the loops  $\al$ are piecewise analytic 
and the gauge group is assumed to be $U(N)$ or $SU(N)$.

However the assumption of analyticity has the unpleasant
consequence that the constructed measures are invariant only under
analytic diffeomorphisms. For the gauge group $U(1)$ a measure was
constructed in \cite{asle} starting from continuous piecewise
$C^k$ loops, for any order $k\geq 1$ of differentiability. In
\cite{baezsawin} the case  of more general gauge group was studied
starting on piecewise smoothly immersed paths.

 The  loop transform  amounts to the construction of a suitable basis  on the Hilbert space of functions on $\bag$ \cite{thiemann}. Other bases are given
in \cite{baezspin} for the analytic case and 
in  \cite{lewandowskithiemann} for the smooth case.
In all these settings the Hilbert spaces under consideration are inductive limits of Hilbert spaces.
Here we treat the Abelian case, $G=U(1)$, where the theory of Abelian groups can be invoked to obtain the loop transform as inductive limit of  Fourier transformations on tori. 

\section{Preliminaries}

We fix a principal fiber bundle $P(M,G)$, where $M$ is an ordinary
 manifold and the group $G$
is  $U(N)$ or $SU(N)$. The manifold $M$ admits a unique
compatible real analytic structure up to $C^{\infty}$
diffeomorphisms (this can be worked out from \S4.7 in \cite{Hirsch}). In the following we shall use a fixed analytic structure.

We will consider  continuous paths and loops on $M$ which are piecewise
analytic, i.e. continuous maps $\gamma$ defined on a closed interval
 $[a,b]$ with a partition $t_0=a\leq t_1\leq...\leq t_i\leq t_{i+1}\leq...\leq t_n=b$ such that
 $\gamma$ agrees with an analytic map
on each $ [t_i,t_{i+1}]$. We call $\gamma$ an {\em
immediate retracing} if 
$ \gamma=\prod_{i=1}^k
\gamma_i\gamma^{-1}_i $
 for a finite collection
$\gamma_1,...,\gamma_k$ of paths in $M$.
 We call a loop $\alpha$ in $M$ {\em thin} if it is homotopic to
the base point $\star$ with an homotopy whose image is entirely
contained in the image of $\alpha$.

Two loops $\al$ and $\be$ are said to be {\em thin-equivalent} if
$\al  \be^{-1}$ is a thin loop.
The thin equivalence agrees with the {\em elementary equivalence},
i.e. equivalence up to (order preserving) reparametrizations and up
to immediate retracings.

The composition of loops based on $\star$ defines a group structure on
 the equivalence classes. We call this group  the {\em group of loops}
and we  denote it by $Loop_{\star}(M)$. For
 sake of simplicity we will denote by $\alpha, \beta ,...$
parametrized loops as
well as their equivalence classes.
\vskip 0,2cm

Let  $\mathcal A$ be the space of smooth connections $A$
on the principal bundle $P(M,G)$ and  ${\mathcal P}_{\alpha}^A
$ the parallel transport defined by $A$ along $\al$. If we fix a
point $u_0$ in the fiber on $\star$, the relation
$u_0={\mathcal P}_{\alpha}^A (u_0)H_A(\al)$ defines a homomorphism
$H_A:\tl\to G$ called  {\em holonomy map} of $A$.

Let $\phi$ be a gauge transformation of $P(M,G)$ (i.e. a $G$-equivariant automorphism of $P$ inducing the identity map on $M$) and $\phi^* A$ be
the pull-back of $A$ by $\phi$. We have
$H_{\phi^*A}(\al)=g_0H_A(\al)g_0^{-1}$, where $g_0$ is the element
of $G$ such that $\phi(u_0)=u_0g_0$.

The Wilson function associated to a loop $\alpha$ is the gauge invariant function defined by $T_{\alpha}(A)=\frac{1}{N}Tr
H_A(\al),$  which can be thought as a bounded complex valued function on $\ag$,
where $\cal G$ denotes the group of gauge transformations.

Taking all the complex linear combinations of finite products
of Wilson functions we get a $\ast$-algebra denoted with
$hol(M,G)$. If we complete $hol(M,G)$ in the $\parallel \;\;\,
\parallel_\infty$ norm we get an Abelian  \cs algebra called  {\em holonomy \cs algebra}, denoted by $Hol(M,G)$. 
If $\star$ is the constant loop,  $T_\star(A)=1$ for every $A$; thus $T_\star$ is the unit $\mathbf I$ in $Hol(M,G)$.
 
We quote here the main results on $Hol(M,G)$ given in several
papers, f.i. \cite{asle} \cite{marolfmourao}; a short review can
be also found in \cite{ln}.

\bi
\item The spectrum of $Hol(M,G)$ is a
compact Hausdorff space in which $\ag$ is densely embedded; for
this reason it is usually indicated with $\bag$; its elements are
called generalized connections and indicated by $\bar A$.

\item $\bag$ agrees with the space $Hom(\tl,G) / Ad_G$ of all homomorphisms
\newline $H:\tl \to G$ up to
conjugation.
\item $Hol(M,G)$ and $hol(M,G)$ do not depend on  the principal bundle
but only on $M$ and $G$.

\ei

\section{A generalization of Bochner Theorem}
  
In this section  the gauge group $G$ is assumed to be $U(1)$ or $ SU(2)$ and we denote by $\cal W$ the set of Wilson functions:
${\cal W}\equiv\{T_{\alpha} \, , \, \alpha \in \tl\} \, .
$
\bd \label{definite} A complex valued function $\ell$ on $\cal W$ is  positive
definite if  $\; \sum_{i=1}^{n}c_{i}\ell(T_{\alpha_i})\geq 0 \;$ whenever
 $\; \sum_{i=1}^{n}c_{i}T_{\alpha_{i}}(\bar {A})\geq 0 \; $ for every $\bar A$.

\ed

We denote by ${\cal B}({\cal W})$ the set of bounded functions and by
${\cal P}({\cal W})$ the set of positive definite functions on $\cal W$; we also identify $Hol(M,G)$ with $C(\bag)$ by means of the Gelfand isomorphism.

Let $M(\bag)$ denote the space of complex regular measures on $\bag$.
To every $\mu\in M(\bag)$, we associate the map $\ell_{\mu}$ on $\cal W$ defined by
 $$
 \ell_{\mu}(T_{\alpha}):=\ibag
 \overline{T_{\alpha}(\bar{A})}d\mu(\bar{A}).$$ 
The function $\ell_{\mu}$ is the restriction to $\cal W$
 of the  bounded linear functional $I_{\mu}$ on $C(\bag)$ 
defined by $I_{\mu}(f)=\ibag f d\mu$ for every $f\in C(\bag)$. We have the following results.

\begin{theorem} \label{theorem:biiezione} 

\begin{enumerate}
\item The map $
  {\cal L} : M(\bag)  \rightarrow  {\cal B}({\cal W}),\quad
  {\cal L}(\mu)=\ell_{\mu} $
 is a continuous injective linear map.

\item The restriction of $\cal L$ to the cone $M_+(\bag)$ of the positive
measures gives rise to a one-to one correspondence with ${\cal
P}({\cal W})$.
\end{enumerate}
\end{theorem}
{\em Proof.} 
First we recall that the algebra $hol(M,G)$ agrees with the
linear span of $\cal W$. This property is obvious for $U(1)$; for  $SU(2)$ it follows 
from the Mandelstam identity:
 $2T_{\alpha}T_{\beta}= T_{\alpha\beta}+
T_{\alpha\beta^{-1}}.$
 
The map $\cal L$ is injective as a consequence of the density of $hol(M,G)$ in $Hol(M,G)$.
The inequality $|\ell_{\mu}(T_{\alpha})|\leq \|\mu\|$ implies  boundeness
of $\ell_{\mu}$ and  continuity of
$\cal L$. Morever, if $\mu$ is a positive measure,  $\ell_{\mu}$ is positive definite.

Let now $\ell \in {\cal P}({\cal W})$. We can associate to $\ell$ a functional $I$ on $hol(M,G)$ defined by
$I(\sum_{i=1}^{n}c_{i}T_{\alpha_{i}})=
\sum_{i=1}^{n}c_{i}\ell(T_{\alpha_{i}}).$
As $\ell$ is positive definite, $I$ is positive 
and well defined: actually 
$
\sum_{i=1}^{n}c_{i}T_{\alpha_{i}}=0\mbox{\, implies \,}\sum_{i=1}^{n}c_{i}
\ell (T_{\alpha_{i}})=0.
$
Now we prove that $I$ is bounded:
for a real valued function $f$ in $hol(M,G)$ we have
$
-\|f\|_{\infty}{\mathbf I}\leq f\leq \|f\|_{\infty} {\mathbf I};
$
positivity of $I$ implies
$|I(f)|\leq\|f\|_{\infty}\ell({\mathbf I}).$
If $f$ is complex valued we obtain the same result by applying the  above argument to the real and imaginary part of $f$.

As $hol(M,G)$ is dense in $Hol(M,G)$ we can extend $I$ to a positive
 continuous
functional $\cal I$ on $Hol(M,G)$ obtaining, by Riesz-Markov Theorem, a regular measure $\mu$ on $\bag$. The positivity of $\cal I$ follows from these considerations: if $f\in Hol(M,G)$, $f\geq 0$, then $\sqrt{f} \in Hol(M,G)$. We can approximate $\sqrt{f}$ with a sequence $p_n$ of elements in $hol(M,G)$. Then
$p_n p_n^*\geq 0$ and $p_np_n^*\to f$ so that ${\cal I}(f)=\lim_nI(p_np_n^*)\geq 0$.
$\qquad\Box$

\vskip 0,2cm
The above results were given in
\cite{asle} with a different definition of positive
definite function.
\vskip 0,2 cm

If $\mu\in M_{+}(\bag)$, then every element $\psi\in  L^{2}(\bag,\mu)$ defines
a measure $\mu_{\psi}$ in $M(\bag)$ such that $\ibag
\overline{T_\alpha} d\mu_\psi = \ibag \overline{T_\alpha} \psi
d\mu $. Hence the loop transform can be obtained simply by restriction of $\cal L$ to $L^{2}(\bag,\mu)$ for a suitable $\mu$. This map is
still an injective linear operator and it is also continuous w.r.t. the $L^2$ norm because $
|\ell_{\mu_\psi}(T_{\alpha})|\leq \|\psi\|_2\|\mu\| ^{\frac{1}{2}}.
$
The range of the loop transform constructed in this way is not
characterized in an explicit fashion. Our subsequent inductive construction will show that, in the Abelian case, an unitary loop transform with  explicitely characterized range is actually available. 
\vskip 0,2cm
We  stress that Theorem \ref{theorem:biiezione} does not depend on
 piecewise analyticity of the loops. Analogous results can be
stated starting on piecewise $C^k$ loops, where $1\leq k\leq
\infty$. In this case $\bag$ will denote the spectrum of the holonomy algebra generated by the related Wilson functions.

For other gauge groups $G$ one can try to generalize Theorem \ref{theorem:biiezione}
using  
 some  subsets $\cal W$ of $hol(M,G)$ containing, besides 
the  Wilson functions, also products of Wilson functions to garantee that their linear span is $hol(M,G)$. Mandelstam identities
can be used to reduce the order of products to be considered.
The  results would be less appealing.

\section{The loop transform in the Abelian  case}
In the following we will fix the gauge group to be $U(1)$. 
We first note that, in this case, $Hoop_\star (M)$ is a group under pointwise multiplication, called the group of hoops in \cite{asle} and denoted by $Hoop_\star(M)$.
We want to prove that $\cal W$ is exactly $\tl$ quotiented
by its commutator subgroup, but to do this we have to introduce the notion of independent loops.

We call {\em edge} in $M$ an
analytic path $ e:[0,1]\rightarrow M $ such that the restriction
$e_{\upharpoonright (0,1)}$ is an  embedding. We call {\em vertex} of an edge the
starting or the ending point.
A finite embedded {\em graph} $\Gamma$ is the image of a finite collection of edges which intersect
themselves only at their vertices.
For any finite collection $\al_1,\ldots,\al_n$ of $n$
(parametrized) loops based on $\star$, the union of their
images ${\al_1}^\ast \cup \ldots \cup {\al_n}^\ast$ is a finite
embedded connected graph $\Ga$. We stress that this result
depends on the piecewise analyticity of the loops.

\begin{definition}\label{independentdef}
The loops $\beta_1,\ldots,\beta_n$ are called { independent}
if every $\beta_i$ contains an edge $e_i$ of the generated graph whose image $e_i^*$ is covered only once by $\beta_i$ and whose image  intersects the other loops only in its vertices.   
\end{definition}

One can prove that { $\beta_1,...,\beta_n$ are independent if and only if
they are the generators of $\pi_1(\Gb)$}.
We recall that $\pi_1(\Gb)$ is free and it can be considered  as a subgroup of $\tl$, so that
an independent family of loops generates a free subgroup of $\tl$
with $n$ generators.

The main property of the independent families of loops are stated in the next proposition (see \cite{asle}).
\begin{proposition}\label{interp} 
The following assertions hold:
\begin{enumerate}
\item every finite family of loops depends on an independent family;
\item let $\beta_1,...,\beta_n$ an independent family  and $(g_1,...,g_n)\in G^n$, then there exists a connection $A\in {\mathcal A}$ such that
$
H_A(\alpha_i)=g_i,\quad i=1,...,n.
$
\end{enumerate}
\end{proposition}
We say that a family of loops depends on another family if every loop of the first family can be written as product of elements of the second one and of their inverses. Property 2. is called the {\em interpolation property}.
\vskip 0,2cm
We introduce the map $\tau: \tl \to {Hoop_\star(M)}$, $\tau(\alpha)=T_{\alpha}$; this map is certainly not injective (owing to the cyclic property of trace, one has $\tau(\alpha)=\tau(\beta\alpha\beta^{-1})$ for any other loop $\beta$). In the Abelian case $\tau$ is a group homomorphism whose kernel  contains the commutator subgroup
of $\tl$, but one can say more.

\begin{proposition}\label{kernel}
Let $G=U(1)$; $\tau$ is a homomorphism whose kernel is the commutator subgroup of $\tl$.
\end{proposition}

{\em Proof.} We recall that the commutator subgroup is the normal subgroup generated by elements of the form $\alpha\beta\alpha^{-1}\beta^{-1}$. By standard algebraic arguments it follows that the elements of the commutator subgroup are of the form:

\begin{equation}\label{ker}
\alpha=\beta_1^{k_{1,1}}\beta_2^{k_{2,1}}...
\beta_n^{k_{n,1}}
\beta_1^{k_{1,2}}...\beta_1^{k_{1,m}}...\beta_n^{k_{n,m}}
\end{equation}
where $\beta_1,...,\beta_n$ are arbitrary elements of the group and $k_{i,j}\in {\mathbb Z}$ satisfy $Q_i \equiv \sum_{j=1}^m k_{i,j}=0 \,$ for every $i=1,...,n$.
Therefore
the commutator subgroup of $\tl$ is  contained in the kernel of $\tau$.

Let $\alpha\in \ker \tau$.
We can write $\alpha$ by means of an independent family of loops 
$\beta_1,...,\beta_n$ as in  formula (\ref{ker}).
Using the interpolation property we can find, for every $\theta\in \reali$ and every $k=1,...,n$, a connection $A_{\theta,k}$ such that:
$
H_{A_{\theta,k}}(\beta_i)=1,\quad \mbox{ for }i\not=k
$
and
$  
H_{A_{\theta,k}}(\beta_k)=e^{i\theta} \, .
$
As $\alpha$ satisfies $T_{\alpha}(A_{\theta,k})=1$ for $\theta \in \reali$ and $k=1,\ldots,n$, we get $Q_k=0$ for every $k$.\qquad $\Box$
\vskip 0,2cm
As a consequence of the identification  $\bag\equiv Hom(\tl,G) / Ad_G$,
we have  that, in the Abelian case,  $\bag$ agrees with the compact Abelian group $Hom(\tl,U(1))$ and $Hoop_\star(M)$ with its dual group. In fact the Wilson functions are normalized characters of $\bag$ so they form an orthonormal set in $L^2(\bag,\mu_0)$, where by $\mu_0$ we denote the Haar measure. This set is complete because the span of the Wilson functions, i.e. $hol(M,U(1))$, is dense in $L^2(\bag,\mu_0)$. By Peter-Weyl Theorem we know that the normalized characters form an orthonormal basis in $L^2(\bag,\mu_0)$. We conclude that ${\cal W} \equiv \widehat{\bag}$. 
\vskip 0,2cm
By a classical result due to A. Weil \cite{weil}, every compact group $\sf G$
is the limit of a projective family of compact Lie groups.
Furthermore, if $\sf G$  is Abelian, it is the limit of a family 
${\sf
G}_\mu$ of compact Abelian Lie groups.
Instead of following Weil's construction, we will use for the compact group 
$Hom(\tl,U(1))$ a more suitable projective family given by Marolf and Mour\~ao in  \cite{marolfmourao}. We will specialize their results for $G=U(1)$
to construct $\bag$ as projective limit of tori and to obtain the loop transform by means of the usual Fourier transforms on tori.

\section{Projective and inductive limits }
We recall the formal definitions and properties of projective and inductive limits
of groups and Hilbert spaces. The definitions we  will give here are not the general ones but adapted  to our situation.

By a  {\em projective family of topological groups}
we mean a collection $ ({\sf G}_{\mu} ,\pi_{\mu\nu} , J)  $ where
 $J$ is a directed set,
  ${\sf G}_\mu$ is a topological group for every $\mu\in J$
 and  the maps $\pi_{\mu\nu}:{\sf G}_\nu\rightarrow {\sf G}_\mu$, defined for every $\mu\leq\nu$,
 are
continuous surjective homomorphisms (called projections) satisfying the consistency conditions:
 \begin{enumerate}
  \item $\pi_{\mu\mu} = id_{{\sf G}_\mu} \, ;$
  \item   $\pi_{\mu \nu}\circ \pi_{\nu \lambda}=\pi_{\mu \lambda}$ for  $\mu \leq \nu \leq \lambda$.
 \end{enumerate}
We call {\em projective limit} of the family
any topological group ${\sf G}$ such that for every $\mu$ there exists a continuous surjective  homomorphism $p_\mu
:{\sf G}\to {\sf G}_\mu$ satisfying  
\begin{enumerate}
 \item $ \pi_{\mu\nu}\circ p_\nu=p_{\mu}$ for $\mu\leq \nu$ ;
\item  $p_\mu(g)=e_{\mu}$ for every $\mu$ implies $g=e$,
\end{enumerate}
where by $e$ and $e_{\mu}$ we denote the units of $\sf G$ and ${\sf G}_{\mu}$, respectively. It is customary to indicate briefly ${\sf G}\equiv
\varprojlim_{\mu} \; {\sf G}_\mu.$  All such topological groups are isomorphic.
\vskip 0,2cm
 If ${\sf G}_\mu$ is a compact  group
for every $\mu\in J$, the projective limit  exists
and it is a compact group.  If ${\sf G}_\mu$ is a connected
for every $\mu$, then  ${\sf G}$ is connected.

\vskip 0.2cm
An {\em inductive family  of topological groups} 
 $(G_\mu,i_{\nu \mu},J)$ is a collection of topological groups $G_{\mu}$,
where $J$ is a directed set of indices and $i_{\nu \mu}:G_\mu \to
G_\nu$ are continuous injective  homomorphisms (called inclusions) defined for every $\nu \geq \mu$ and satisfying the consistency conditions:
\ben
 \item $i_{\mu \mu}=id_{G_\mu}$ ;
 \item $i_{\lambda \nu} \circ i_{\nu \mu} = i_{\lambda \mu}
 $ for $ \mu \leq \nu \leq \lambda$.
\een

We call {\em inductive limit} of the family any topological group $G$ 
such that for every $\mu$ there exists a continuous
injective homomorphism $i_{\mu}:G_{\mu}\to G$ satisfying: 
\begin{enumerate}
\item $
i_{\nu}\circ i_{\nu\mu}=i_{\mu}\quad \mu\leq \nu;
$
\item the entire $G$ is covered by  the union of the images of the inclusions $i_{\mu}$.  
\end{enumerate}

We indicate briefly $G=\lim_{\mu}G_{\mu}$.
All such topological groups are isomorphic.
\vskip 0,2cm
The dual group  $\hat{\sf G}$ of  a locally compact Abelian group $\sf G$ which is a projective limit of a family of locally compact Abelian groups  $G_{\mu}$  is the inductive limit  of the dual groups $\hat{\sf G}_{\mu}$ with the transposed maps $^ti_{\mu}$ and $^ti_{\nu\mu}$ as projections; this result can easy worked out by \S 5 of
A.Weil's book \cite{weil}. 

\vskip 0,2cm 

We specialize the definition of inductive family and inductive limit  to
 the category of Hilbert spaces by requiring the inclusions to be isometric linear maps.
In this case we define the inductive limit to be any Hilbert space $\hc$ such that  the inclusions $i_{\mu}$ cover a dense linear subspace of  $\hc$. All such Hilbert spaces are isomorphic.

 Our definition of  inductive family of Hilbert spaces is quite restrictive.
In fact every inductive family  $(\hc_\mu,i_{\nu \mu},J)$
 generates a projective family  $(\hc_\mu,\pi_{ \mu \nu},J)$, taking as projections $\pi_{ \mu \nu}$ the adjoint maps 
$^t i_{\nu \mu}$. The inductive limit $\hc$ with projections $\pi_{\mu}=^ti_{\mu}$ 
is also the projective limit of the family $(\hc_\mu,\pi_{\mu \nu},J)$. 

\vskip 0,2cm

Finally we give the definition of {\em inductive and projective limit of unitary maps}.
Let $\{ \hc_\mu,i_{\nu \mu},J \}$ and $\{ {\hc '}_\mu, {i'}_{\nu \mu}, J \}$ be  inductive families of Hilbert spaces, with inductive limits $\hc$ and $\hc'$, respectively. A family of   unitary maps $\{ \phi_\mu : \hc_\mu \to {\hc '}_\mu\}_{(\mu \in J)}$ is said to be inductive if it satisfies
 $$ {i'}_{\nu \mu} \circ \phi_\mu=\phi_{\nu} \circ i_{\nu \mu}\quad \mbox{for}\quad\mu \leq \nu.$$
 The inductive limit is the unique  unitary map $\phi : \hc \to\hc'$ satisfying 
$$\phi \circ i_\mu = {i'}_{\mu} \circ \phi_\mu\quad\mbox { for every }\quad \mu.$$

Analogously, in the projective case, a family of unitary maps $\{\phi_\mu\}_{(\mu\in J)}$ is a projective family if it satisfies
 $$ {\pi'}_{\mu \nu} \circ \phi_\nu= \phi_\mu \circ \pi_{\mu \nu} \quad \mbox{ for }\quad \mu \leq \nu.$$
Their projective limit is the unique unitary map $\phi : \hc \to\hc'$ such that

$$ \pi'_\mu\circ\phi =  \phi_\mu\circ\pi_\mu\quad\mbox{ for every }\quad \mu.$$

\section{Inductive construction of the loop transform in the Abelian case}\label{inductiveconst}
We first construct the group $\bag$ as a projective limit of tori following \cite{marolfmourao}.  
Let us consider the set $J$ of the subgroups $L$ of $\tl$ generated by a finite independent family of loops. By $L \leq L'$ we mean that $L$ is a subgroup of
$L'$; $J$ is  directed  w.r.t. this ordering.

The projective family associated to $\bag$ is defined as follows:

\begin{itemize}
\item we take as index set the directed set $J$;
\item to every $L \in J$ we associate the group $ Hom(L,U(1))$; 
\item if $L \leq L'$ we define the projection $\pi_{L L'}:
Hom(L',U(1))\to Hom(L,U(1)) $ which restricts  the homomorphisms $H \in Hom(L',U(1))$ to the subgroup $L$.
\end{itemize}
To simplify the notation we denote $Hom(L,U(1))$ by $\bag_L$ and its dual group  by ${\cal W}_L$. 
 For a given independent family of loops $(\alpha_1,...,\alpha_n)$
the evaluation map 
\begin{equation}\label{evaluation}
ev_{(\alpha_1,...,\alpha_n)}(H)=(H(\al_1),...,H(\al_n))
\end{equation}
is an isomorphism of $\bag_L$ with the n-dimensional torus.

The group $\bag \equiv Hom(\tl,U(1))$ is  the projective limit of this family; actually the projection $\pi_l:\bag\to\bag_L$, wich restricts the homomorphisms $H\in Hom(\tl,U(1))$ to $L$, is continuous and surjective owing to the interpolation property of independent loops. \vskip 0,2cm

The loop transform $\cal L$ will be constructed as the inductive limit of the
Fourier transforms $F_{L}$ between the Hilbert spaces $L^2(\bag_L)$
and $L^2({\cal W}_L)$ (for shortness we have omitted the
relative Haar measures).
The scheme of the work is visualized in this diagram:
\begin{displaymath}
\begin{array}{ccc}
  \vdots & \vdots & \vdots \\
  L^2(\bag_{L}) & \stackrel{F_L}{\longrightarrow} & L^2({\cal W}_L) \\
  \downarrow i_{L'L} & \vdots & \downarrow j_{L'L} \\
  L^2(\bag_{L'}) & \stackrel{F_{L'}}{\longrightarrow} & L^2({\cal W}_{L'}) \\
  \vdots & \vdots & \vdots \\
  \downarrow & \downarrow & \downarrow \\
  L^2(\bag) & \stackrel{\cal L}\longrightarrow & L^2({\cal W})\equiv L^2(Hoop_\star(M))
\end{array}
\end{displaymath}

\vskip 0,2cm
To make the family $\{L^2(\bag_{L})\}$ an inductive family of
Hilbert spaces we define the  inclusions $i_{L'L}$ for 
 $L \leq L'$ as follows: for every $\psi \in L^2(\bag_{L})$ we put
  $$(i_{L' L}\psi)(H'):=\psi(\pi_{L L'}(H')) \qquad H'\in \bag_{L'}. $$

These inclusions are linear and satisfy the
consistency conditions, so we  have only to prove that they are
isometric maps. 
Suppose that $L$ and $L'$ are the free groups generated by the independent families $\{ \al_1,\ldots,\al_n\}$ and $\{\be_1,\ldots,\be_{n'}\}$, respectively, and that $L \leq L'$. We have:
\begin{displaymath}
 \left\{ \begin{array}{l}
  \al_1={\be_1}^{k_{1,1}}  \ldots  {\be_{n'}}^{k_{n',1}} \\
  \vdots \\
  \al_n={\be_1}^{k_{1,n}}  \ldots  {\be_{n'}}^{k_{n',n}}
  \end{array} \right.
\end{displaymath}
for some $k_{r,s}\in \mathbb{Z}$ for $r=1,\ldots,n'$ and $s=1,\ldots,n$.

For
$ev_{(\beta_1,\ldots,\beta_{n'})}(H)=(e^{i\vartheta_1},\ldots,e^{i\vartheta_{n'}}),
$ it follows  $$ev_{( \alpha_1,\ldots,\alpha_n )}(\pi_{L' L}(H))=
(e^{ik_{1,1}\vartheta_1} \, \cdots \,
e^{ik_{n',1}\vartheta_{n'} },\ldots,e^{ik_{1,n}\vartheta_1} \, \cdots
\, e^{ik_{n',n}\vartheta_{n'}}) \, . $$
By composition of the evaluation maps with $i_{L'L}$ one obtains the inclusions
 $i_{n'n} : L^2(U(1)^{n}) \to L^2(U(1)^{n'})$, defined by $$(i_{n'n}\psi)
(e^{i\vartheta_1},\ldots,e^{i\vartheta_{n'} }) =
\psi(e^{i(k_{1,1}\vartheta_1 +...+k_{n',1}\vartheta_{n'}
)},\ldots,e^{i(k_{1,n}\vartheta_1+...+k_{n',n}\vartheta_{n'})}).$$

From the normalization and the bi-invariance of the Haar 
measure it follows that the inclusions $i_{n'n}$, and hence also the inclusions $i_{L' L}$,
are isometric.

The inclusions $j_{L' L}$ are defined by the following commutative diagram:
\begin{displaymath}
\begin{array}{ccl}
 L^2(\bag_{L'}) & \stackrel {F_{L'}}{\longrightarrow} & L^2({\cal W}_{L'}) \\
  \uparrow _{i_{L'L}}       &                                     & \uparrow  _{j_{L' L}}    \\
 L^2(\bag_{L}) & \stackrel {F_L}{\longrightarrow}    & L^2({\cal W}_L)
\end{array}
\end{displaymath}
They are isometries as compositions of isometric maps. The diagram shows that the consistency conditions hold both for the inclusions $j_{L' L}$ and the Fourier transforms $F_L$. So we have well defined inductive families.

\begin{theorem} 

\ben
\item $L^2(\bag)$ is the inductive limit of $\{ L^2(\bag_{L}) \}_L$;
\item $L^2({\cal W})$ is the inductive limit of $\{ L^2({\cal W}_L) \}_L $; 
\item the loop transform $\cal L$ on $L^2(\bag)$ is the inductive limit of $\{ F_L \}_L$.
\een
\end{theorem}

{\em Proof.}  Let us define  the inclusions $i_L :L^2(\bag_L)\to L^2(\bag)$ by $(i_L \psi_L)(H)=\psi_L(\pi_L(H))$, $\psi_L \in L^2(\bag_L)$.

Denoting by $\mu_L$ the Haar measure on $\bag_L$ and by $\mu_0$ the normalized Haar measure on $\bag$,  we have that $(\pi_L)_*\mu_0=\mu_L$. Therefore 
$$ \| i_L \psi_L\|^2=\ibag \mid \psi_L \circ \pi_L \mid^2
d\mu_0(\bar A) = \int_{\bag_L} \mid \psi_{L} \mid^2
d\mu_{L} = \| \psi_L \|^2 \; $$
so that the inclusions $i_L$ are isometric.
Moreover their images  contain the Wilson functions, hence they cover a dense linear subspace of $L^2(\bag)$. We conclude that $L^2(\bag)$ is the inductive limit of the family $\{L^2(\bag_L)\}_L$.

Now we define the inclusions $j_L:L^2({\cal W}_L)\to
L^2({Hoop_\star(M)})$ by the following commutative diagram: 
\begin{displaymath}
\begin{array}{ccl}
 L^2(\bag)     & \stackrel {\cal L}{\longrightarrow}   & L^2({Hoop_\star(M)}) \\
 \uparrow _{i_{L}}   &                                  & \uparrow  _{j_{L}}    \\
 L^2(\bag_{L})        & \stackrel {F_L}{\longrightarrow} & L^2({\cal W}_L)
\end{array}
\end{displaymath}
Repeating the same arguments on the inclusions $j_L$ we get that  $L^2({Hoop_\star(M)})$ is the inductive limit of the family $\{L^2({\cal W}_L)\}_L$ and that the inductive limit of $\{ F_L \}_L$ is the loop transform on $L^2(\bag)$. $\qquad \Box$

\vskip 0,2cm
From ${\cal L} \circ i_L = j_L \circ F_L$ we get the transposed equality $p_L \circ F^{-1} = {F_L}^{-1} \circ \hat{p}_L$, where  $p_L:= {^t i_L}$ and $\hat{p}_L:= {^t j_L}$. This shows that the inverse Fourier transforms $F_L^{-1}$ form a projective family of unitary maps whose limit is the inverse loop transform.
The inverse loop transform of a ``loop state'' $\ell: {Hoop_\star(M)}\to \complessi$  is given by
$\psi=\sum_{\alpha\in Hoop_\star(M)}\ell(\alpha)T_{\alpha}.$

\section{The smooth case}

A piecewise smooth  path is a continuous map $p$ on a closed interval $[a,b]$ in $M$ with a finite partition $t_0=a\leq t_1\leq ...\leq t_i\leq t_{i+1}\leq ...\leq t_n=b$ such that $p$ agrees with a $C^{\infty}$ curve in every  interval $[t_i, t_{i+1}]$.
Again we can contruct  the group of equivalence classes of piecewise smooth  loops based on $\star$ up to reparametrizations and retracings; we will denote it by
  ${\sf Loop}_\star(M)$.  
We recall that ${\sf Loop}_\star(M)$ is a topological group  endowed with  the  topology generated by piecewise smooth homotopies, i.e. curves in ${\sf Loop}_\star(M)$ defined by continuous maps
$\Phi:\reali\times [0,1]\to M$ which are  smooth on  $\reali\times [t_k,t_{k+1}]$
 for some partition $t_0=0<t_1<...<t_n=1$. This topology was introduced by Barrett
  in \cite{barrett}.

For a given $U(1)$-bundle $P$ on $M$, the  group of hoops, denoted by ${\sf Hoop}_\star(M)$, is the
quotient of ${\sf Loop}_\star(M)$ with respect to  the subgroup  
 $ \{\alpha\in {\sf Loop}_\star(M)\;|\;H_{A}(\alpha)=1 \quad\forall A\}$,
where $A$ denotes a  connection on  $P$. We will call this subgroup  {\em holonomy kernel}.   The following proposition follows  from the characterization of the hoops  given in \cite{spallanzani}.
\begin{proposition}
The holonomy kernel is the closure of the commutator subgroup in the Barrett topology.
\end{proposition}
{\em Proof:}
The holonomy kernel contains the closure of the commutator subgroup: in fact  
the holonomy maps $H_A$ associated to  connections $A$ are  continuous in the Barrett topology and so  the holonomy  kernel is closed in this topology. 
Following Proposition 5.8 in \cite{spallanzani} every loop in the holonomy kernel can be approximated by a loop in the commutator subgroup.
 $\qquad\Box$
\vskip 0,2cm
 
As a consequence of this characterization the hoop group is an Abelian group 
 not depending on the bundle $P$ and can be constructed using the trivial bundle. 
Moreover ${\sf Hoop}_\star(M)$ is torsion free (see Lemma A.2 in \cite{asle}).

In a torsion free $\mathbb Z$-modulus every finitely generated submodulus is freely generated;  then in every finite generated subgroup $L$ of ${\sf Hoop}_\star(M)$ we can
 choose 
a finite family  $\tilde {\alpha_1}, ..., \tilde{\alpha_n}$  of free generators
and give an isomorphism of the group $Hom(L,U(1))$ with $U(1)^n$ as in \S  \ref{inductiveconst}.

In the case of  trivial bundles  and of  bundles arising as pullback of the Hopf bundle $S^1\to S^3\to S^2$  the following  weak form of the interpolation property holds, which assures 
that the spectrum of the holonomy algebra in the smooth case agrees with the compact Abelian group $Hom({\sf Hoop}_\star(M),U(1))$ (see \cite{asle}).

\begin{proposition}\label{inter}
For a family $\tilde {\alpha_1}, ..., \tilde{\alpha_n}$  of free generators the
 evaluation map
$
ev_{\alpha_1,...,\alpha_n}(A)=(H_A (\alpha_1),...,H_A(\alpha_n)) 
$
has range dense in $U(1)^n$. 
\end{proposition}

Then a situation arises  analogous to the analytic case.
\begin{proposition}
 $Hom({\sf Hoop}_\star(M),U(1))$ is the projective limit of the family of Abelian compact groups $\{Hom(L,U(1))\}_L$ where the index $L$ runs over the finitely generated subgroups of ${\sf Hoop}_\star(M)$.
\end{proposition}

{\em Proof:} The only non trivial point is to show that the continuous projections
$p_L:Hom({\sf Hoop}_\star(M),U(1))\to Hom(L,U(1))$ are surjective. This follows by proposition \ref{inter} using the fact that the image of $p_L$ must be compact.
$\qquad \Box$

\vskip 0,2cm
As in the analytic case we get that the group ${\sf Hoop}_\star(M)$ is the dual group of \newline
$Hom({\sf Hoop}_\star(M),U(1))$ 
 and that a construction of the loop transform as an inductive limit of Fourier transforms of tori can be performed also in the smooth case.
The remarkable difference is that a set of independent hoops is not easy characterized  as in the analytic case.

\section{The path transform}
 Let $G$ be a compact group, $A$ denote a connection on the trivial bundle $M\times G$ and $F:G^k\to \mathbb C$ a continuous function and $p_1,...,p_k$ piecewise analytic  paths in $M$; we can define 
the function $f$ on $\cal A$ by 
by
$$
f(A)=F(H_A(p_1),...,H_A(p_k)),
$$
where $H_A(p)$ denotes the parallel transport along $p$ defined by the connection $A$,  identified  with an element of $G$.
Functions of this form are called {\em cylinder functions} and are contained in ${\cal B}({\cal A})$.
They generate a $C^*$-algebra with unit,  called briefly {\em the cylinder $C^*$-algebra}, whose spectrum $\overline{\cal A}$ contains  $\cal A$ densely.

The space $ \pt $ of the equivalence classes of piecewise analytic paths $p$ in $M$ up to reparametrizations and retracings is a groupoid where the composition $p_1p_2$  is defined  if the end point of $p_1$ agrees with the starting point of $p_2$ and the inverse $p^{-1}$ is obtained by reversing the parametrization. Every parallel transport
 $H_A:\pt\to G$ is a groupoid homomorphism:   
$H_A(p_1p_2)=H_A(p_1)H_A(p_2)$.

One can define families of independent paths as in Definition \ref{independentdef}. Every family of paths depends on a family of independent paths and  the interpolation property holds  for independent paths as stated in Proposition \ref{interp} for independent loops. It follows that $\overline {\cal A}$
 agrees with 
the space $Hom(\pt,G)$ of all homomorphisms from $\pt$ to $G$.
The proof of this result is similar to the one used in \cite{asle} to prove that
$\bag=Hom(\tl,G)/Ad G$ (see also \cite{velhinho}).
$Hom(\pt,G)$ is a closed subset  of the compact group $G^{\pt}$. 

In the Abelian case  $\overline {\cal A}=Hom(\pt,U(1))$ is an Abelian compact group and
it is the projective limit of the family of the compact groups $ Hom(L,U(1))$
where $L$ is the subgroupoid generated by a finite family of independent paths.
Actually the interpolation property assures that the projections $\pi_L:Hom(\pt,U(1))\to Hom (L,U(1)$, defined by restrictions, are surjective  homomorphisms. 

\begin{proposition}
1) The dual group $\widehat{Hom}(\pt,U(1))$ is  generated by the  maps
$\chi_p:{Hom(\pt,U(1))}\to U(1), \;\chi_p(H)=H(p).$

2) The kernel of the homomorphism $\tau:{ \pt}\to\widehat{Hom}(\pt,U(1))$, $\tau(p)=\chi_p$ is the subgroupoid generated by elements of the form
\begin{equation}\label{path}
p_1^{k_{1,1}}p_2^{k_{2,1}}...p_n^{k_{n,1}}p_1^{k_{1,2}}...
p^{k_{1,m}}_1...p_n^{k_{n,m}}
\end{equation}
where $k_{i,j}\in \mathbb Z$ and $Q_i=\sum_{j=1}^m k_{i,j}=0$
 \end{proposition}

{\em Proof.}
Every $\chi_p$ is a continuous character: it is multiplicative  and  continuous as restriction to $Hom(\pt,U(1))$ of the projection $\pi_p:U(1)^{\pt}\to U(1)$
 on the $p$ component.
The group $X$ generated by the characters $\chi_p$ is separating on $Hom(\pt,U(1)$. Then $X$ is  the entire dual group (apply Theorem 23.20 in \cite{hewitt}).

2) It follows by the interpolation property as in Proposition \ref{kernel}.
$\qquad\Box$

\vskip 0,2cm
In the Abelian case one can introduce the path transform which is simply  the Fourier transform on $\overline{\cal A}$.  An inductive construction of the path transform can be obtained, as in \S\ref{inductiveconst} in the case of the loop transform,  using as index set the family of subgroupoids generated by independent paths.

\end{document}